\documentclass[aps,pra,longbibliography,preprint,amsmath,
amssymb,floatfix,superscriptaddress]{revtex4-2}

\usepackage{graphicx}
\usepackage{dcolumn}
\usepackage{bm}
\usepackage{hyperref}
\usepackage[mathlines]{lineno}
\usepackage{natbib}
\usepackage[T1]{fontenc}
\usepackage[english]{babel}
\usepackage{amsthm}
\usepackage{tabularx}
\usepackage{float}
\usepackage{tikz}
\usepackage{tikz-cd}

\newcommand{\Ec}{\mathcal{E}}

\newcommand{\Bbf}{\mathbf{B}}

\newcommand{\Vbf}{\mathbf{V}}

\newcommand{\nbf}{\mathbf{n}}

\newcommand{\zbf}{\mathbf{z}}

\newcommand{\SO}{\mathrm{SO}}

\newcommand{\fict}{\mathrm{fict}}
\newcommand{\ficts}{\mathrm{fict}^{(1)}}

\newcommand{\ii}{\mathrm{i}}

\newcommand{\dd}{\mathrm{d}}
\newcommand{\TT}{\mathrm{T}}

\theoremstyle{plain}

\newtheorem*{thm*}{Theorem}
\newtheorem*{lem*}{Lemma}
\newtheorem*{prop*}{Proposition}
\newtheorem*{cor*}{Corollary}

\theoremstyle{definition}

\begin{document}
\title{Optically Induced Monopoles, Knots, and Skyrmions in Quantum Gases}
\author{Toni Annala}\email{tannala@ias.edu}\affiliation{School of Mathematics, Institute for Advanced Study, 1 Einstein Drive, Princeton, 08540, NJ, USA}\affiliation{QCD Labs, QTF Centre of Excellence, Department of Applied Physics, Aalto University, P.O. Box 13500, FI-00076 Aalto, Finland}
\author{Tommi Mikkonen}\affiliation{Photonics Laboratory, Physics Unit, Tampere University, P.O. Box 692, FI-33014 Tampere, Finland}\affiliation{Department of Chemistry, University of Helsinki, P.O. Box 55, FI-00014 Helsinki, Finland}
\author{Mikko M{\"o}tt{\"o}nen}\affiliation{QCD Labs, QTF Centre of Excellence, Department of Applied Physics, Aalto University, P.O. Box 13500, FI-00076 Aalto, Finland}
\date{\today}

\begin{abstract}
We propose, and theoretically analyze, a practical protocol for the creation of topological monopole configurations, quantum knots, and skyrmions in Bose--Einstein condensates by employing fictitious magnetic fields induced by the interaction of the atomic cloud with coherent light fields. It is observed that a single coherent field is not enough for this purpose, but instead we find incoherent superpositions of several coherent fields that introduce topological point charges. We numerically estimate the experimentally achievable strengths and gradients of the induced fictitious magnetic fields and find them to be adjustable at will to several orders of magnitude greater than those of the physical magnetic fields employed in previous experimental studies. This property together with ultrafast control of the optical fields paves the way for advanced engineering of topological defects in quantum gases. 
\end{abstract}

\maketitle

Topology is the mathematical theory of the qualitative properties of shapes~\cite{milnor:1965,hatcher:2002}. Its roots can be traced back to the origins of graph theory~\cite{bondy:1976}, and, in its modern form, to an 1895 article by Poincare~\cite{poincare:1895}. In physics, topological structures have attracted persistent attention ever since the vortex atom hypothesis of Kelvin in 1869~\cite{thomson:1869}, suggesting atoms to be knotted vortex loops in the ubiquitous ether. Even though this hypothesis was incorrect, an abundance of other topological structures has been discovered since then. Examples in fundamental physics include the Dirac monopole~\cite{dirac:1931} and cosmic strings~\cite{bucher:1992, bucher:1992b, bucher:1994}. Moreover, condensed-matter systems support a wide variety of topological phenomena, such as topological defects~\cite{volovik:1977, mermin:1979} in concrete physical systems such as superfluids~\cite{volovik:2009}, liquid crystals~\cite{machon:2013, beller:2014, machon:2019}, and Bose--Einstein condensates~\cite{kawaguchi:2012, ueda:2014}, the quantum Hall effect~\cite{thouless:1982, thouless:1983}, and topological phases of matter, such as topological insulators~\cite{kane:2005, kane:2005b, moore:2007, fu:2007, hasan:2010} and topological superconductors~\cite{read:2000, kitaev:2001, sato:2017}. Accordingly, topological phenomena in physics has developed into a vibrant research area in which experiments~\cite{ray:2014, ray:2015, hall:2016, lee:2018, blinova:2023} and simulations~\cite{ruostekoski:2003, faddeev:1997, priezjev:2002, kartashov:2014,driben:2014} meet the methods of abstract mathematics~\cite{annala-mottonen, annala:2022, annala-rajamaki-mottonen, rajamaki:2023}, and which has a wide range of potential applications to, for example, high-temperature superconductivity~\cite{peotta:2015} and fault-tolerant quantum-computation~\cite{kitaev:2003}.

Ultracold quantum gases and especially atomic Bose--Einstein condensates (BECs) with spin degree of freedom~\cite{kawaguchi:2012} provide a highly controllable system where the coherent quantum state of the gas can be imaged with high resolution. Thus they seem ideal for studying the topological properties of matter. Based on a theoretical proposal~\cite{pietila:2009b} that the topological winding of the direction of a three-dimensional quadrupole magnetic field can be used to imprint Dirac monopoles in BECs, these intrigiung defects were observed for the first time in any continuous field~\cite{ray:2014}, followed by first creations of topological monopoles~\cite{ray:2015}, quantum knots~\cite{hall:2016}, three-dimensional skyrmions~\cite{lee:2018}, and Alice rings~\cite{blinova:2023}. However, a practical protocol for creating multiple defects inside a single cloud---a prerequisite for studying defect interactions---has been essentially lacking. A key challenge in manufacturing such configurations has been the difficulty of creating magnetic fields that vary in a sufficiently nontrivial fashion inside the small spatial extent of a Bose--Einstein condensate cloud by employing a conventional magnet~\cite{tiurev:2019}.

A promising avenue for bypassing this issue is the manipulation of the internal degrees of freedom of atoms and molecules by coherent light~\cite{suter:1997}, which is one of the most versatile tools in the toolbox of the modern experimental physicist. The traditional applications of this phenomenon include magneto-optical traps and laser cooling~\cite{raab:1987}, which paved the way to the first experimental observation of BECs~\cite{anderson:1995} and their topological vortex defects~\cite{matthews:1999}. Moreover, optical driving fields have been employed in the creation of two-dimensional skyrmions in spin-2 BECs~\cite{wright:2008, leslie:2009}, and several protocols based on optical methods have been proposed for generating skyrmions and possibly knotted vortex-loop configurations in Bose--Einstein condesates \cite{ruostekoski:2001,ruostekoski:2003,ruostekoski:2005,parmee:2022}. In addition, fine spatial control of the trapping potential allows the fabrication of optical lattices, enabling the simulation of a variety of quantum systems~\cite{gross:2017}. 

In this paper, we are focusing on the creation of topologically winding fictitious magnetic fields generated by atom-light interactions, suitable for creating a variety of topological defects in BECs, including monopoles, quantum knots, three-dimensional skyrmions, and configurations thereof. The effect of a coherent, off-resonant light field can be described by an effective Hamiltonian~\cite{cohen-tannoudji:1972}, which has three interaction terms, the \emph{scalar}, \emph{vector}, and \emph{tensor light shift}. The scalar light shift usually plays the predominant role in applications. However, the effects of the vector light shift, equivalent to the effects of a \emph{fictitious magnetic field} $\Bbf_\fict$, cannot be overlooked, for example, in the estimation of the transition energies relevant for atomic clocks~\cite{flambaum:2008, rosenbusch:2009}. Importantly, large-gradient fictitious fields are a versatile tool~\cite{albrecht:2016} that have been used in cooling atoms below the Doppler limit~\cite{perrin:1998, hamann:1998, kerman:2000}, in building deformable optical lattices~\cite{sebby-strabley:2006} which are useful for the realization of two-qubit gates~\cite{anderlini:2007, lundblad:2009} in neutral-atom quantum computing, and in optical nanofiber traps~\cite{vetsch:2010, kien:2013b, schneeweiss:2014}. The tensor part of the light shift is negligible for the ground states of alkali metal atoms, and does play a role in our investigations.  


Let us consider the possibility of creating point-defect configurations in BECs by employing fictitious magnetic fields, thus addressing a key challenge in the case of complex point-defect configurations. Previously, single monopoles have been imprinted in BECs by adiabatically moving a topologically nontrivial magnetic field inside the cloud by slowly ramping a homogeneous bias magnetic field~\cite{pietila:2009b, ray:2014, ray:2015}. Unfortunately, the extension of this scheme to simultaneously create several point defects is experimentally challenging due to the strong magnetic fields required to create the required topologically nontrivial field~\cite{tiurev:2019}. To resolve this issue, we propose to replace the physical magnetic field by a fictitious field $\Bbf_\fict$, the shape, strength, and stability of which are determined by the light sources used to induce $\Bbf_\fict$. We find that the large gradients attainable~\cite{albrecht:2016} offer an opportunity for tailoring the micrometer-scale shape of the fictitious field, which is a prerequisite for creating a small enough point-defect configuration to fit inside the spatial extent of a BEC cloud. Thanks to the established experimental methods to control optical fields with high precision and resolution in space and time, our proposal uncovers paths to conquer new regimes for the experimental creation of various topological configurations.

However, our observations suggest that there is a subtlety inherent to this strategy: the fictitious magnetic field $\Bbf_\fict$ generated by a single coherent laser field contains no topological charges, and therefore cannot alone be employed to create even a single point defect. Nonetheless, we demonstrate that it is possible to obtain topologically non-trivial fictitious fields by considering an incoherent superposition of multiple laser fields, the fictitious field of which is the sum of the fictitious fields of the individual components~\cite{cohen-tannoudji:1972, goldman:2014}. Consequently, we arrive below at an advanced technique for experimentally creating monopole and other topological configurations in spinor BECs.

Let us consider an atom of total hyperfine angular momentum $F$ interacting with a single coherent light field which is characterized by the complex-valued polarization vector $\Ec$. The corresponding fictitious magnetic field experienced by the atom is given by~\cite{rosenbusch:2009, kien:2013}
\begin{equation}\label{eq:fict}
\Bbf_\ficts := \frac{\alpha_v}{8 g_F \mu_B F} \ii ( \Ec^* \times \Ec ), 
\end{equation} 
where $\alpha_v$ is the vector polarizability of the atom, $g_F$ is the Landé~$g$ factor, and $\mu_B$ is the Bohr magneton. Thus, a spatially varying polarization vector may lead to a spatially varying fictitious magnetic field. Denoting $\Ec = \Vbf + \ii \mathbf{W}$, where $\Vbf$ and $\mathbf{W}$ are three-dimensional real-valued vector fields, we observe that
\begin{equation}\label{eq:fict2}
\Bbf_\ficts \propto \Vbf \times \mathbf{W}.
\end{equation}

\begin{figure}[h!]
\includegraphics[scale=1.4]{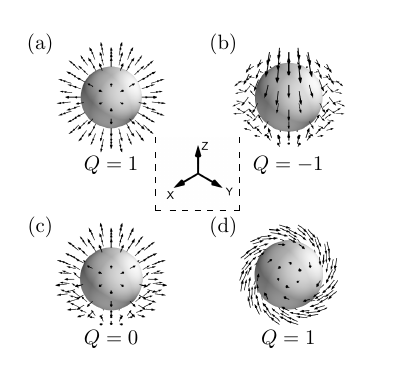}
\caption{
\textbf{Examples of topological charges.}
(a) Hedgehog configuration containing a topological charge $Q = 1$ at the origin. Reversing the directions of the arrows results in the anti-hedgehog configuration, which contains a topological charge $Q = -1$.
(b) Vector field that has the shape of a quadrupole field $[x,y,-2z]^\TT$ contains topological charge $Q = -1$ at the origin, since a global rotation of the vectors by an angle of $\pi$ about the $z$ axis results in essentially the anti-hedgehog configuration.
(c) Vector field not attaining all possible directions leads to a vanishing topological charge.
(d) Vector field of the form $[y,z,x]^\TT$ contains topological charge $Q=1$ at the origin, since a cyclic permutation of the coordinate axes has no effect on $Q_\Omega$ in Eq.~\eqref{eq:topocharge}. 
}\label{fig:topcharge}
\end{figure}

As we show below, a fictitious field $\Bbf_\ficts$ of Eq.~\eqref{eq:fict2} does not contain any topologically non-trivial winding along any sphere which implies that the field does not contain any point-like \emph{topological charges}, also referred to as topological monopoles, examples of which are illustrated in Fig.~\ref{fig:topcharge}. 
More formally, the number of times the field winds along an arbitrary sphere $\Omega$, on which the fictitious field obtains only nonzero values, can be obtained from
\begin{equation}\label{eq:topocharge}
Q_\Omega := \frac{1}{8 \pi} \int_{\Omega} \dd\omega_i \epsilon_{ijk} \hat\Bbf_\ficts \cdot \bigg(\frac{\partial \hat\Bbf_\ficts}{\partial x_j} \times \frac{\partial \hat\Bbf_\ficts}{\partial x_k} \bigg) = 0,
\end{equation}
where  $\epsilon_{ijk}$ is the fully anti-symmetric tensor introducing an implicit summation over $i, j, k\in\{1, 2, 3\}$, the coordinate system is given by $(x_1,x_2,x_3)=(x,y,z)$, and the hat on the vectors denotes unit norm. We mathematically consider $Q_\Omega$ as the \emph{degree}~\cite{milnor:1965} of the induced map $ \hat\Bbf_\ficts |_\Omega: \Omega \to S^2$, where we have identified the two-dimensional sphere $S^2$ with the space of three-dimensional unit-vectors. The equality $Q_\Omega = 0$ holds if and only if $\hat\Bbf_\ficts |_\Omega$ is \emph{nullhomotopic}, i.e., homotopic to a constant map~\cite{milnor:1965}. In other words, a field that has no topological winding, can be continuously transformed without passing through zero into a homogeneous configuration.

To show the nullhomotopy that $Q_\Omega = 0$ for all $\Bbf_\ficts \propto \Vbf \times \mathbf{W}$, we observe that the normalized ficticious field can be expressed as
\begin{equation}
\hat\Bbf_\ficts = \Vbf' \times \mathbf{W}',
\end{equation}
where $\Vbf'$ and $\mathbf{W}'$ are obtained from $\Vbf$ and $\mathbf{W}$ on $\Omega$ by applying the Gran--Schmidt orthonormalization process. The triad $(\Vbf', \mathbf{W}', \hat\Bbf_\ficts)$ may be regarded as a continuous map $\Omega \to \SO(3)$, and since any such map is null-homotopic~\cite{annala-mottonen}, we obtain the desired null-homotopy of $\hat\Bbf_\ficts$ from the null-homotopy of the triad. Thus, the fictitious field induced by a single, coherent light field contains no topological point charges.

To overcome the above obstruction, we incoherently superimpose several coherent light fields, generated by several independent laser sources. The fictitious field induced by such a superposition is the sum of the fictitious fields generated by the individual coherent fields~\cite{cohen-tannoudji:1972}. 
Alternatively, one could employ several lasers of slightly different wavelengths, in such a way that none of the frequency-differences are close to the resonance frequencies of the atom. This ensures that the temporal dependence of the fictitious fields induced by the frequency offsets can be neglected in the dynamics of the condensate~\cite{goldman:2014}. 

\begin{figure}[h!]
\includegraphics[scale=.85]{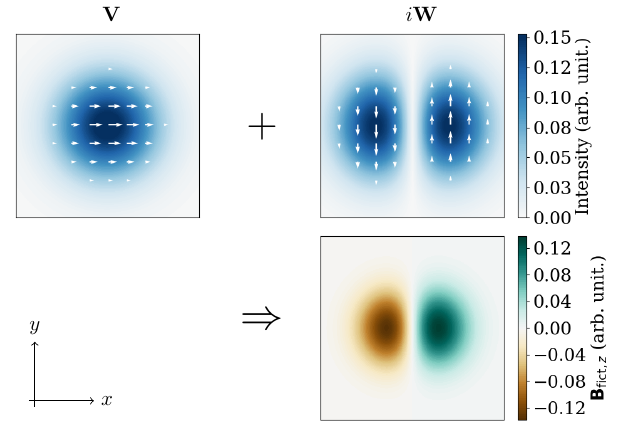}
\caption{
\textbf{Fictitious field generated by a coherent superposition of a Gaussian and a Hermite--Gaussian beam.}
By coherently superimposing a Gaussian $\mathrm{TEM}_{00}$ beam (top left panel) and a Hermite--Gaussian $\mathrm{TEM}_{10}$ beam (top right panel) that have orthogonal cross-sectional linear polarization fields $\Vbf$ and $\mathbf{W}$, respectively, and a $\pi/2$-phase-difference,
results in a fictitious field $\Bbf_\fict$ (bottom panel) that  aligns with the propagation direction ($z$-axis) of the beam. The top panels indicate the beam intensities and linear polarization directions of $\Vbf$ and $\mathbf{W}$. The bottom panel indicates the $z$-component of the fictitious field.
}\label{fig:thebeam}
\end{figure}

The cross-sectional polarization vector of a Hermite--Gaussian $\mathrm{TEM}_{mn}$ beam at focus, propagating along the $z$ axis and polarized along a vector $\hat \nbf$  in the $xy$ plane, is given by
\begin{equation}\label{eq:polarizationv}
    \Ec_{m,n}(x,y) = A_{m,n} \exp\left(-\dfrac{x^2 + y^2}{w^2_0}\right) H_m\left( \dfrac{\sqrt{2} x}{w_0} \right) H_n\left( \dfrac{\sqrt{2} y}{w_0} \right) \hat \nbf
\end{equation}
where parameter $w_0$ is the \emph{waist} of the beam~\cite{erikson:1994} and the constant $A$ is given by
\begin{equation} \label{eq:Econstant}
    A_{m,n} = \sqrt{\frac{4P}{\pi w_0^2\epsilon_0c_02^{m+n}m!n!}},
\end{equation}
where $P$ is the optical power of the beam, $\epsilon_0$ is the permittivity of vacuum, and $c_0$ is the speed of light in vacuum. Hence, a coherent superposition of a $\mathrm{TEM}_{00}$ and a $\mathrm{TEM}_{10}$ beam, with orthogonal polarization and a $\pi/2$ phase difference, induces for $x,y\ll w_0$ a fictitious field of the form
\begin{equation}
\Bbf_{\fict}^{(00,10,z)} \propto 
x \hat\zbf,
\end{equation}
as we illustrate in Fig.~\ref{fig:thebeam}. 
Adding three such fields generated by incoherently superimposing pairs of beams propagating along $x$, $y$, and $z$ directions respectively, we obtain the total fictitious field 
\begin{equation}\label{eq:Bfict-tot}
\Bbf^\mathrm{tot}_{\fict}=\Bbf_{\fict}^{(00,10,x)}+\Bbf_{\fict}^{(00,10,y)}+\Bbf_{\fict}^{(00,10,z)} \propto 
\begin{bmatrix}
y & z & x
\end{bmatrix}^\TT,
\end{equation}
that contains a topological charge $1$ at the origin as shown in Fig.~\ref{fig:topcharge}(d).

\begin{figure*}[h!]
\includegraphics[scale=0.45]{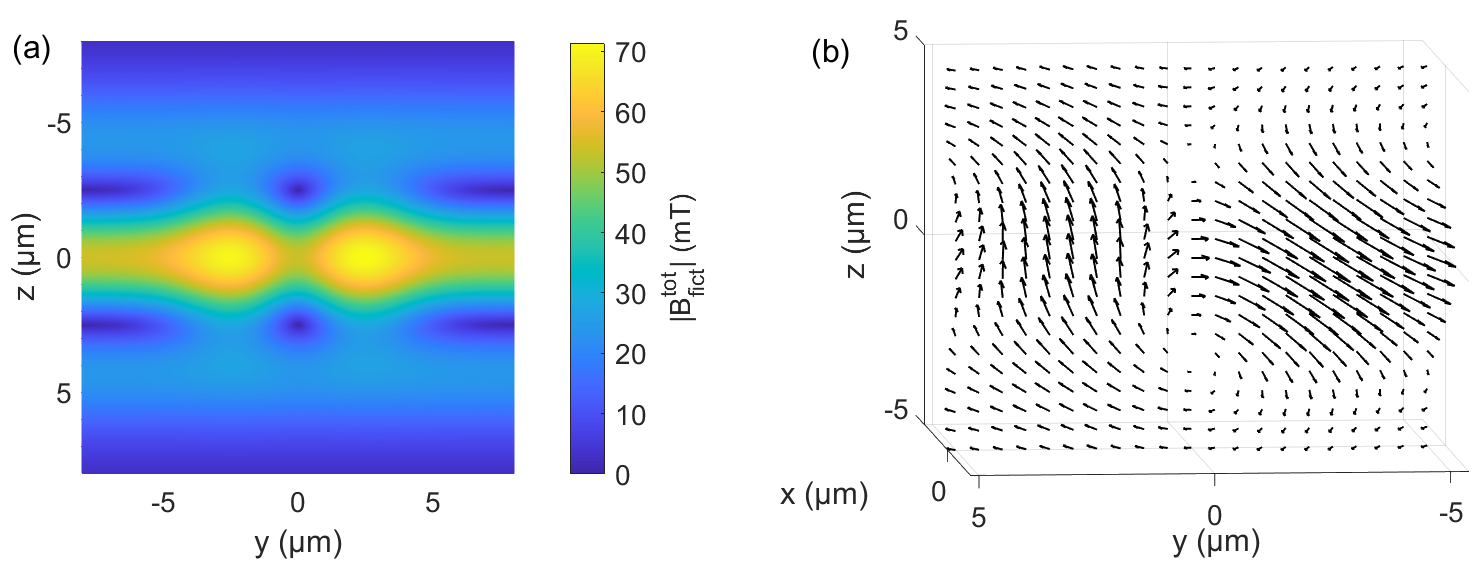}
\caption{\textbf{Fictitious magnetic field containing a topological monopole-antimonopole pair.}
(a) Strength of the total fictitious magnetic field in the $x=0$ plane. The field is induced by an incoherent superposition of three pairs of coherent light beams propagating along $x$, $y$, and $z$, respectively. The beams propagating along $x$ and $z$ are coherent superpositions of Hermite--Gaussian $\mathrm{TEM}_{00}$ and $\mathrm{TEM}_{10}$ beams. The beam propagating along $y$ is a coherent superposition of a $\mathrm{TEM}_{00}$ and a  $\mathrm{TEM}_{20}$ beam. The zero points of the field are roughly 5~µm apart. Each of the six beams has a wavelength of 770~nm, a beam waist of 5~µm, and 1~W of optical power. The material parameters were $\alpha_v = 4462$~a.u.~\cite{arora:2012} ($7.357\times10^{-38}$~A$^2$s$^4$kg$^{-1}$), $F=1$, and $g_F=-1/2$. (b) Direction of the field shown in (a) in the $x=0$ plane indicated by the arrows.
}\label{fig:Btot}
\end{figure*}

In order to create more complicated point-defect configurations, one or several of the above $\mathrm{TEM}_{10}$ beams can be replaced by a higher-order Hermite--Gaussian beam $\mathrm{TEM}_{n0}$ with $n>1$. For instance, if one $\mathrm{TEM}_{20}$ beam is used together with two $\mathrm{TEM}_{10}$ beams, the total fictitious field $\Bbf^\mathrm{pair}_{\fict}=\Bbf_{\fict}^{(00,10,x)}+\Bbf_{\fict}^{(00,20,y)}+\Bbf_{\fict}^{(00,10,z)}$ contains a topological monopole-antimonopole pair in the vicinity of the origin as shown in Fig.~\ref{fig:Btot}. In general, any configuration of monopoles and antimonopoles in a three-dimensional charge-alternating grid may be generated. 

The effects of a physical and a fictitious magnetic fields combine additively~\cite{goldman:2014}. Hence, a monopole configuration may be imprinted inside the sample by first producing a fictitious field with the corresponding point-defect configuration, and then adiabatically moving the zero point(s) of the fictitious field inside the cloud in order for the order parameter to keep aligned with the field, by slowly ramping down the strength of a physical bias magnetic field as in Refs.~\cite{ray:2014, ray:2015}. Similarly, a configuration of either quantum knots or three-dimensional skyrmions may be created by rapidly switching on the fictitious field, inducing position-dependent spin rotations that form topological solitons with nontrivial $\pi_3$-charges as in Refs.~\cite{hall:2016,lee:2018}. To capture the full topology of the fictitious field, its zero-point configuration needs to be small enough to fit inside the BEC cloud. 

Alternatively to the physical magnetic field, one may employ two phase shifted $\mathrm{TEM}_{00}$ beams to induce an additional essentially homogeneous fictitious field at the BEC and move the zero points by changing the intensity of the fields. In case there are physical off-set magnetic fields to cancel, one may need homogeneous fictitious fields in three linearly independent directions. This kind of all-optical control of the magnetic fields enables orders of magnitude faster field ramps than those with physical magnetic fields, which can be advantageous for example in creating quantum knots~\cite{hall:2016} and skyrmions~\cite{lee:2018} where the zero point of the field is instantaneously brought into the condensate.

Next, we describe the parameter values for the experimental creation of fictitious magnetic fields containing topological monopole-antimonopole pairs that are small enough to fit inside a typical BEC that is 15~µm in diameter, similar to Ref.~\cite{tiurev:2019}. We consider a $^{87}\mathrm{Rb}$ BEC with $\alpha_v = 4462$~a.u.~\cite{arora:2012} ($7.357\times10^{-38}$~A$^2$s$^4$kg$^{-1}$), $F=1$, and $g_F=-1/2$, but parameter values for other atomic species may be obtained in a similar manner.
We consider the beam configuration of Eq.~\eqref{eq:Bfict-tot} and Fig.~\ref{fig:thebeam}, where all beams have a waist of $w_0 = 5$~µm, a wavelength of $\lambda = 770$~nm, and optical power of $P=1$~W.

There are several approaches to create a coherent superposition of Hermite--Gaussian beams of different orders. For example, one can split a Gaussian beam and transform one of the beams into a higher order mode using phase plates~\cite{meyrath:2005}, or employ two phase-locked lasers with intra-cavity elements to control the transverse profiles of each laser beam. Alternatively, two different Hermite--Gaussian beams with orthogonal polarizations may be produced in the same cavity using a birefringent beam displacer or a polarizing beam splitter to partially separate the beam paths in the cavity and phase elements to provide gain only for the selected transverse modes~\cite[pages 359--362]{oron:2001}. An additional phase plate can be added into one of the paths for maintaining coherence with a selected phase difference between the modes. Regardless of the approach, the resulting beams are focused in order to obtain the desired beam waist.


The three pairs of coherent beams will be focused on the BEC with orthogonal propagation directions. 
Since the distance between the two zero-points of a $\mathrm{TEM}_{20}$ beam at focus coincides with the waist, we expect the total fictitious field $\Bbf^\mathrm{pair}_{\fict}=\Bbf_{\fict}^{(00,10,x)}+\Bbf_{\fict}^{(00,20,y)}+\Bbf_{\fict}^{(00,10,z)}$ to contain a topological monopole and an antimonopole that are five microns apart~\cite{siegman:1986}. 
To confirm the position of the field zeros and to estimate the strength of the total fictitious field $\Bbf^\mathrm{pair}_{\fict}$, we calculate fields $\Bbf_{\fict}^{(00,10,x)}$, $\Bbf_{\fict}^{(00,20,y)}$, and $\Bbf_{\fict}^{(00,10,z)}$ using Eqs.~\eqref{eq:fict},~\eqref{eq:polarizationv}, and~\eqref{eq:Econstant} in a three-dimensional grid around the origin.
For each pair of coherently superimposed Hermite--Gaussian laser beams, we evaluate Eq.~\eqref{eq:fict} across the transverse plane at the BEC using the known electric-field profiles of linearly polarized Hermite--Gaussian modes~\cite{erikson:1994}. In addition to the dominant polarization component considered in Eq.~\eqref{eq:polarizationv}, we also include cross and longitudinal polarization components that may become significant when strongly focusing the beam~\cite{erikson:1994}, but here their contribution is rather small. Since the considered spatial lengths ($\sim$10~µm) are an order of magnitude shorter than the Rayleigh range of the beam, we can neglect changes in the transverse profile and the Gouy phase of the laser beams during propagation, and assume for each beam that the magnetic field from Eq.~\eqref{eq:fict} remains constant along the direction of beam propagation.  

The sum of the three fields yields the total fictitious field $\Bbf^\mathrm{pair}_{\fict}$ shown in Fig. \ref{fig:Btot}. The position of the field zeros is approximately 5~µm as expected. The maximum strength of the fictitious field is 71~mT, and the spatial gradient field between the monopole-antimonopole pair is about 22~kT/m. This gradient is almost six orders of magnitude greater than roughly 40~mT/m used in previous monopole experiments~\cite{ray:2014,ray:2015}. Thus by adjusting the power of the lasers, it is convenient to span a very broad range of gradients.

At $100$~\textmu{W} optical power, the maxima of the scalar light shifts caused by the coherent superpositions $\mathrm{TEM}_{00} + \mathrm{TEM}_{10}$ and $\mathrm{TEM}_{00} + \mathrm{TEM}_{20}$, are roughly $10^{-28}$~J each. 
Thus, with $5$~µm waist, the total effect of the incoherent superposition employed in the defect creation corresponds to a repulsive $100$-Hz trap, which is weaker than the optical traps previously employed in studying monopoles in BECs. Moreover, we estimate the ground-state-loss of atoms owing to the spontaneous scattering of the trapping laser to be of the order of $10$~$\mathrm{s}^{-1}$~\cite{pethick:2008}, corresponding to BEC lifetime of the order of $100$~ms. In contrast, the defect creation time of roughly 0.5~ms has been previously used for quantum knots and skyrmions\cite{hall:2016,lee:2018}, and 40~ms for monopoles\cite{ray:2014,ray:2015,blinova:2023}. Thus with $100$~\textmu{W} optical power, corresponding to 2.2-T/m gradient, the spurious effects of the scalar light shift and spontaneous scattering are not dominating even if the defect creation time and optical trap is taken as in previous experiments, but we obtain two-orders-of-magnitude improvement in the gradient field.

We described a practical protocol for the imprinting topological monopole, quantum-knot, and skyrmion configurations in Bose--Einstein condensates by employing fictitious magnetic fields induced by the ac Stark Shift. Using experimentally feasible parameters, the gradients of the generated fictitious fields seem orders of magnitude greater than previously obtained by physical magnetic fields. Combining this result with those of previous numerical simulations~\cite{tiurev:2019}, the proposed protocol seems feasible for creating, for example, a Dirac monopole-antimonopole pair in the ferromagnetic phase of spin-1 BEC, and a topological monopole-antimonopole pair in the polar phase of spin-1 BEC~\cite{vuojamo:MSc}. The experimental implementation of the protocol is an appealing future direction of research.

Our analysis demonstrates the versatility of the fictitious fields generated by light-matter interactions. In particular, fictitious fields seem to have the advantage of convenient shaping of micrometer-scale textures compared to physical magnetic fields. They can also offer an all-optical scheme to extremely quickly control the fictitious magnetic fields free of typical sources of electrical noise in coils, thus opening opportunities for unprecedented accuracy in creating quantum knots, skyrmons, and possibly new type of topological textures. We estimate that 1-W laser beams may be used to create a quantum knot in less than a nanosecond. Future studies may reveal whether it is possible to extent our configuration of point-like field zeros to line-like zeros with non-trivial topological consequences.


\section*{Acknowledgments}
The authors would like thank Roberto Zamora-Zamora, Joonas Vuojamo, and David Hall for useful discussions. We have received funding from the Research Council of Finland Centre of Excellence program (project nos. 352925 and 336810), from the Vilho, Yrj\"o and Kalle V\"ais\"al\"a Foundation of the Finnish Academy of Science and Letters, from the National Science Foundation
(Grant No. DMS-1926686), from the Fulbright Finland Foundation, and from the Emil Aaltonen Foundation. T.A. thanks the IAS for excellent working conditions.

\clearpage 

\clearpage

\bibliography{references}

\end{document}